# Crafting a systematic literature review on open-source platforms

Jose Teixeira and Abayomi Baiyere

TUCS - Turku Centre for Computer Science
University of Turku
Finland
Email: jose.teixeira@utu.fi, abayomi.baiyere@utu.fi

**Abstract**. This working paper unveils the crafting of a systematic literature review on open-source platforms. The high-competitive mobile devices market, where several players such as Apple, Google, Nokia and Microsoft run a platforms-war with constant shifts in their technological strategies, is gaining increasing attention from scholars. It matters, then, to review previous literature on past platforms-wars, such as the ones from the PC and game-console industries, and assess its implications to the current mobile devices platforms-war. The paper starts by justifying the purpose and rationale behind this literature review on open-source platforms. The concepts of open-source software and computer-based platforms were then discussed both individually and in unison, in order to clarify the core-concept of "open-source platform" that guides this literature review. The detailed design of the employed methodological strategy is then presented as the central part of this paper. The paper concludes with preliminary findings organizing previous literature on open-source platforms for the purpose of guiding future research in this area.

# 1   Introduction

## 1.1   Purpose and rationale

The mobile devices market has been extremely competitive within the last five years. Apple, Google, Nokia and Microsoft among others played a very dynamic platforms-war, seeking control over the distribution of software and content to mobile hardware devices such as smartphones, netbooks and computer-tablets. The open-source software plays an important role in this platforms-war. As an indication - Apple reveals that open-source is a key part of its ongoing software strategy [1] and Google claims to lead the development of the Android platform by open-source approach [2]. On other hand, Nokia decided to give-up open-source software by closing down Symbian and Meego [3] and adopting Microsoft Windows Phone for its smart-phone strategy [4]. Yet another player; Hewlett-Packard, made big shifts on its technological strategy by abandoning WebOS, a mobile platform also based in open-source software components, after investing millions on its development[5].

An increasing number of researchers within the Information Systems (IS) field have addressed the ongoing mobile platforms-war from multiple perspectives. Mian et al. reported some implications of the open-source phenomenon on the ongoing platforms-war by studying the technological strategies employed by Apple, Google and Nokia [6]. From an innovation studies perspective, Eaton et al. explored the paradoxical relationship between control and generativity of innovation in digital ecosystem by having Apple and Google as units of analysis [7]. Building on th boundary objects theory and innovations networks literature, Ghazawneh and Henfridsson developed a process perspective of third-party development governance through boundary resources by studying the Apple's iPhone developer program [8]. From software architecture and licensing perspectives, Anvaari and Jansen evaluated the architectural openness of five different mobile platforms concluding that Google's Android and Nokia's Symbian were the most open platforms [9].

Evidently, the mobile platforms-war is gaining attention from the IS research community. However behind the *vogue*, it is important to assess how this emergent mobile platforms-war is different from previous platforms-wars covered by previous decades of published literature. This raises the following questions: Is the literature from previous platforms-wars, such as in PC and the game-console industries addressing this current war between Apple, Android, Microsoft and others?

For addressing this and other questions, we decided to execute a systematic literature review on open-source platforms, embracing a need for more and better documented literature reviews on the IS field[10]. This paper addresses the call from von Brocke et al. for the publication of two versions of the same literature review [11]. One that contains all the major findings, to be published later; and another that outlines the literature search process. This current paper addresses the latter. Subsequently, we discuss the concepts of open-source software and computer-based platforms followed by the employed methodology based on established guidelines on how to conduct a systematic literature review in the IS field.

### 1.2   On the evolving open-source phenomenon

There is a consensus of four freedoms expressed by Stallman [12] which laid the foundation of the open-source phenomenon [12]:

- *The freedom to run the program, for any purpose.*
- *The freedom to study how the program works and change it so it does your computing as you wish.*
- *The freedom to redistribute copies so you can help your neighbor.*
- *The freedom to distribute copies of your modified versions to others.*

For the good of the open-source community, the Open Source Initiative (OSI) was founded by Bruce Perens and Eric Raymond in 1998 to develop and maintain a more commonly agreed open-source definition, based on the social contract from the Debian Linux distribution [13]. Moreover, the OSI open-source definition introduced a novel connection between open-source software and standards [14].

According to Perens, open-source concerns not only software source code but also the distribution terms of software, as visible in the previous FSF and OSI free and open-source software definitions [15]. Both Stallman's and OSI definitions address very well the public with both expertise in software development and software license agreements, however general public could reveal difficulties in understanding the open-source term.

To position the open-source software concept used in this review with a mapping of Stallman's and OSI definitions, we propose three open-source criteria. First, the blue-print availability (software source-code is available upon request); Second, explicit intellectual propriety licenses not restricting users free-software freedoms (Software license empowers user rights); and thirdly, the compliance with standards (the software privileges the use of standards that enable interoperability).

### 1.3   On computer-based platforms

The platform term is conceptually abstract and is widely used across many fields. Within this research, the platform term maps the concept of computer-based platform as previous addressed by Morris, Ferguson, Bresnahan, Greenstein and West [16]–[18]. As argued by West [18], platform consists of an architecture of related standards, controlled by one or more sponsoring firms [18]. The architectural standards typically encompass a processor, operating system (OS), associated peripherals, middleware, applications, etc. Platforms can be seen as systems of technologies that combine core components with complementary products and services habitually made by a variety of firms (complementors). Jointly the platform leader and its complementors form an "ecosystem" for innovation, that increases platform's value and it consequent users' adoption [19]

For example, the once leading Japanese video games industry, operate by developing the hardware consoles and its peripherals while providing a programmable software platform that allows others to develop games on top of their systems. The attraction of more game developers to the platform means more games and an increase of value for the final users (video game players). High-tech firms competing in a high-networked economy must adopt platform based strategies versus product based strategies, due to the difficulty of satisfying an increasing complex consumer demand [20]. In the development of certain complex systems, an "all in house" strategy might not be economically feasible, organizations must adopt "platform-thinking" and focus efforts on the highest value-adding components of the platform, making it open and attractive to all possible participants.

Within this research, the authors address literature on open-source computer-based platforms: meaning computer-based platforms that not only integrate open-source software components, but also provide a set of publicly available open-source components. Prominent examples can be the Google's Android, Apple iOS and Nokia Maemo platforms empowering mobile devices. For instance, all the vendors integrate the WebKit open-source web browser engine into their platforms while providing their modified WebKit versions in open-source manners. It is important to note that, within this reviews context, computer-based platforms combine hardware and software but can also be pure-software platforms. Krishnamurthy and Tripathi [22] and Teixeira [23] studied platforms structured over pure software artefacts. Platforms leaders provide and set the boundaries of their technological-core and provide additional development mechanisms that allow third-parties to complement while adding value to the overall platform under network effects.

## 2     Research methodology and design

After clarifying the core-concepts of "open-source software" and "computer-based platforms" we present the methodology used for the review in this section.

### 2.1    Research goals and methodological base

Primarily and most importantly, by conducting this structured literature review the authors aim to provide an aggregated vision of what is well known within the academia regarding open-source platforms. The underlying research questions are:

- **RQ1:** What are the seminal works bridging open source and platforms?
- **RQ2:** Does the literature from previous platforms-wars, such as in the PC and the game-console industries, address this current mobile-platforms war between Apple, Android, Microsoft and others?
- **RQ3:** What is the seminal literature to be taken into account by researchers and practitioners addressing the ongoing mobile platforms-war?
- **RQ4:** Which previous research findings can't be generalized for such novel and contemporary scenario?

This review considers methodological guidelines provided by Webster and Watson [10], Järvinen [23], von Brocke et. al. [11] and Okoli and Schabram [24]. Transparency and rigour in documenting the literature review process, the use of a systematic and future reproducible procedure; were some of the base-pillars of this review. Simple and common available software tools, like spreadsheet software (LibreOffice), citation manager (Zotero), graph visualization software (Graphviz) and a mind-mapping tool (Xmind) eased the literature review process.

The literature review process started in November 2010, the final set of articles were retrieved on March 2011 and were carefully read and analyzed while taking in account the different methodological guidelines on conducting a literature review.

### 2.2   Design and research basis

After reading the literature review guides and analyzing a small set of systematic review articles published in the IS field, the authors decided to follow closely the literature review design from von Brocke and Theresa [25]. As in [25], the authors made use of Emerald, EBSCO and ProQuest ABI/Inform databases of general journals and conferences; plus the use of Google books index on published books; and finally the use of the eLibrary system from the Association of Information Systems (AIS) as a database indexing more specific journals and conferences within IS field.

The authors decided to complement von Brocke and Theresa research basis by including the Volter national database that indexes books within a large national libraries network. The following Table 1 summaries the database sources from where the literature was collected. All databases were accessed using authors host University Internet proxy even if accessed remotely.

**Table 1.** Database sources used for conducting the literature review.

| Source | Type | Website http:// |
|---|---|---|
| Emerald | General journals and conferences | emeraldinsight.com |
| EBSCO | General journals and conferences | web.ebscohost.com |
| ProQuest ABI/ Inform | General journals and conferences | search.proquest.com |
| Google books | Published books | books.google.com |
| Volter database | Published books | volter.linneanet.fi |
| AIS eLibrary | IS journals and conferences | aisel.aisnet.org |

For retrieving literature that aggregates both knowledge on open-source and computer-based platforms, previous knowledge of the authors reinforced by discussions within the academic circle influenced the choice of the keywords for the search.   Addressing the open-source term, the keywords "open source", "open-source", "OSS", "FLOSS" and "libre" were used. Moreover, for capturing relevant literature within computer-based platforms the keywords "platform", "platforms", "platform-based", "eco system", "eco systems", "eco-system" and "eco-systems" were employed. The decision to use several keywords increased the amount of relevant literature included in the review.

The authors discarded publications that were not relevant to the IS field after a careful content analysis.  Most of the publications discarded did not fit with this papers adopted definitions of open-source and platforms. The authors documented and tabulated each discarded publication item while building associated exclusion criteria. The search was limited to peer-reviewed publications; several published books and journal articles were discarded because they did not clearly meet this criterion. The search was also limited to research expressed in the English-language.

The research basis ($\alpha$) was defined by searching, within the mentioned source databases, for publication items with both open-source and platforms keywords on their titles. An initial set of fifteen publications were defined as the starting point for our research. The fifteen publications included books, two conference proceedings and the remaining were serial journals. After an extensive analysis of the research basis, the authors decided to extend the research ($\beta$) by searching for articles with keywords capturing open-source on their titles and with keywords capturing platforms on the abstract. A total of 360 new publications were identified with this first research extension. For future research, the authors consider the possibility of extending the research to include other publications with platforms on the title and open-source on the abstract.

The following Table 2 gives an overview on how the captured research publications were retrieved by each of the six source databases. A considerable number of collisions, publications indexed more by different databases, was encountered.  Books and dissertation databases were not considered in our research extension because books databases do not support queries addressing a possible book abstract.

**Table 2.** Number of captured research publications per database source.

|  | EMEE | EBS | PQA | GOB | VOL | IAS |
|---|---|---|---|---|---|---|
| $\alpha$ (title CONTAINS (open-source AND platforms)) | 0 | 3 | 5 | 6 | 0 | 1 |
| $\beta$ ((title CONTAINS open-source) AND (abstract CONTAINS platforms) | 1 | 24 | 318 | QNS | QNS | 2 |
| Total captured items per database source | 1 | 27 | 323 | 6 | 0 | 3 |

### 2.3 Extraction and categorization of literature

In order to provide both a quantitative and qualitative overview of relevant research of open-source platforms within the Informations Systems field, the authors extracted and categorized the literatures according to their meta-description and content. The authors first conducted a simpler categorization of the literature without looking at its full content. Some of the retrieved articles were discarded by its meta-description (i.e. after reading the abstract). After reading each articles meta-description, the authors moved afterwards to a more demanding phase, where the deep reading of each papers content enabled the extraction and categorization of research on open-source platforms.

For the first step, content independent information was extracted and categorized by using meta-descriptions of each captured paper. Not all non-content information was available within the used sources databases, requiring visits to the different publishers Internet resources. For each paper found, a manual citation analysis was made using both the http://scholar.google.com and the http://www.isiknowledge.com web resources. The authors decided to keep track of each captured research paper price, if applicable: both the payment amount charged by the publisher to download the paper and the yearly subscription rate, all for later arguing on the cost of this literature review.

For the second step, and in order to provide a qualitative overview of the literature review, the authors delved into the articles content. This started an ongoing demanding analysis of each captured paper, identifying key information such as research questions, methodology, outlined future research, research propositions, theoretical implications, implications for practice, key references, among other content information. After full paper reading and using spreadsheets, the authors systematically retrieved for each paper, information about the research questions being addressed; their triggers and motivations, implications for theory and practice, methodology and philosophical standings, perceived theoretical and empirical relevance, etc. For the very specific context of this literature review the authors also captured for each item what are the research related industry verticals and platforms being studied. For each paper, the authors complemented the collected information in a spreadsheet with two to three slides containing the message of each paper,

After the content analysis, the authors made the transition from author to concept-centric approach as suggested by Webster and Watson [10]. A long concept matrix was developed for mapping the analyzed publication items with key concepts that emerged during the literature review process, e.g. the concepts of "Community of Practice" [26] and "Sense of Community" [27]. Relationships between these key concepts were then mapped using diagram tools (i.e Graphviz and Xmind) providing a theoretical overview[1] of previous research in open-source platforms.

---

[1] Theoretical overview within Gregor's nature of theory in information systems research [28]

## 3    Preliminary findings

As previously mentioned, this literature review is still a work in progress. So far, the meta-description analysis of the retrieved 360 articles is completed. However; just 170 of the articles has been fully read and content-analyzed. This literature review is aimed to be systematic, rigorous and exhaustive which turned out to be a slow process lasting several years. In this section, we present our preliminary findings by revisiting the initial research questions and outlining future research.

**3.1 Revisiting the research questions**

The first research question was "What are the seminal works on open-source platforms?". Based on a citation analysis of the retrieved publications on Google and Thomson Reuters services; and by its recurrence within the articles analyzed so far, the authors proposes: The economic works of Economides and Katsamakas[29]; the open-source adoption studies of Dedrick and West[14]; [30] and the R&D management strategy work of West[18]; as seminal works on open-source platforms.

Our second research question inquired "if research addressing the current mobile-platforms takes in consideration literature from previous platforms-wars?" The third and related initial research question is "What is the seminal literature to be taken in account by researchers and practitioners addressing the ongoing mobile platforms-war?" After reviewing ad-hoc emergent literature on the novel mobile-platforms war such as: Basole's visualization in a converging mobile ecosystem[31]; Eaton et al. description of the paradoxical relationship between control and generativity on Apple and Google ecosystems[7] ; or the innovation study from Remneland-Wikhamn et al. on the iPhone and Android mobile platforms; we claim that emergent research addressing the current mobile-platforms is not considering, or exploiting previous seminal works on open-source platforms, as it often should.

Out last initial research question inquired if previous research findings, on previous platforms-wars, can be generalized to the current mobile platforms-war, scenario. Previous seminal works from Economides, Katsamakas, Dedrick and West [14], [18], [29], [30] assume a scenario where open-source is an alternative strategy for low-cost players, with reduced market-share, against more successful corporations enjoying a quasi-monopoly situation. Using researchers own words:

> *"When a system based on an open source platform with an independent proprietary application competes with a proprietary system, the proprietary system is likely to dominate the open source platform industry both in terms of market share and profitability. This may explain the dominance of Microsoft in the market for PC operating systems."* in [29]

> *"On the other hand, Microsoft's proprietary platform strategies continued to be successful"* in [18]

> *"The most important driver of adoption was cost "* in [14]

*"The major factors are cost, perceived reliability, compatibility ..."* in [30]

Tables turned: First of all, open-source is no longer associated with low-cost products within the current mobile platforms-war. Moreover, the traditional proprietary software players, such as Microsoft and Blackberry, are currently struggling with residual sales on the mobile devices market [32]. Apple, Google and Google Android partners are effectively dominating the market, while charging more for their high-end devices than their competitors[33], all with strategies that esteem open-source software[1], [2].

### 3.2 Future research

When contrasting previous literature on older "platforms-wars", such as the ones from the PC and game-console industries, with the current and under-studied mobile platforms-war, we empirically notice that many of the market players remain the same (Microsoft and Apple). There is a scenario of convergence: same firms push for similar technological standards across different platforms, i.e. Microsoft Windows within X-box, Surface Tablets, PC, Netbooks and Mobile phones. This convergence between industries remains unexplored by academia. Interesting research questions dealing with the implications of such convergence remain unexplored, i.e "should firms concentrate on one platform-war or run several platform-wars in parallel?